\documentclass[apjpt4]{aastex}
\def\chandra{{\it Chandra}}
\def\xmm{{\it XMM-Newton}}
\def\FeKa{Fe K$\alpha$\ }
\begin{document}
\title{The Difference in Narrow \FeKa Line Emission Between Seyfert 1 and Seyfert 2 Galaxies}
\author{Teng Liu \& Jun-Xian Wang}
\affil{CAS Key Laboratory for Research in Galaxies and Cosmology, Department of Astronomy, University of Science and Technology of China; lewtonstein@gmail.com; jxw@ustc.edu.cn}
\begin{abstract}
	We compile a sample of 89 Seyfert galaxies with both [OIV] 25.89 $\mu$m line luminosities observed by Spitzer IRS and X-ray spectra observed by \xmm\ EPIC.
Using [OIV] emission as proxy of AGN intrinsic luminosity, we find that although type 2 AGNs have higher line equivalent width,
the narrow \FeKa line in Compton-Thin and Compton-Thick Seyfert 2 galaxies are  $2.9^{+0.8}_{-0.6}$ and $5.6^{+1.9}_{-1.4}$ times weaker  in terms of luminosity than Seyfert 1 galaxies respectively.
This indicates different correction factors need to be applied for various types of AGNs before the narrow \FeKa line  luminosity could serve as intrinsic AGN luminosity indicator. 
We also find Seyfert 1 galaxies in our sample have on average marginally larger line width and higher line centroid energy, suggesting contamination from highly ionized Fe line or broader line emission from much smaller radius, but this effect is too weak to explain the large difference in narrow  \FeKa line luminosity between type 1 and type 2 AGNs.
This is the first  observational evidence showing the narrow \FeKa line emission in AGNs is anisotropic. The observed difference is consistent with theoretical calculations assuming a smoothly distributed obscuring torus, and could provide independent constraints on the clumpiness of the torus.
\end{abstract}
\keywords{ galaxies: active -- galaxies: nuclei -- galaxies: Seyfert -- X-rays: galaxies -- line: profiles }
\section{Introduction}
Narrow \FeKa emission line cores are  common in the X-ray spectra of both type 1 and type 2 Active Galactic Nuclei (AGNs) (\citealp[e.g. see][]{bianchi09, lamassa09} for most recent studies). 
Such narrow lines have been traditionally associated with an origin in distant matter, especially the putative obscuring torus. Comprehensive and systematic studies of the narrow \FeKa line in Seyfert 1 galaxies with the \chandra\ High Energy Grating (HEG) were presented by \citet{yaqoob04} and \citet{shu10}.
Shu et al. reported a mean line width of 2060 km/s, supporting an origin in distant matter. Meanwhile, considering the observed larger scatter in line width from source to source, contributions from inner region (such as the outer part of the accretion disk) could not be ruled out. Although the velocity widths in some of the sources were found consistent with those of the optical broad emission lines \citep[also see ][]{yaqoob01, bianchi03}, \citet{nandra06} found no correlation between the \FeKa core width and the BLR (specifically H$\beta$) line width, eliminating the BLR as a general origin of the \FeKa core (also see Shu et al. 2010).
The lack of  correlation between the equivalent width of narrow \FeKa emission and EW(CIV) \citep{wu09} gives weight to this scenario.

Type 2 AGNs are believed viewed at inclination larger than type 1 AGN in the framework of unification model \citep[e.g.,][]{antonucci93}. In such model, the line of sight of the observer to type 2 AGNs is obscured by an equatorial torus, which blocks the central accretion disk and the broad emission line region. The column density of the obscuration in type 2 AGNs spans from $N_H \sim 10^{22} cm^{-2}$ to $N_H > 10^{24} cm^{-2}$ (Compton-Thick) \citep{bassani99}.
The narrow \FeKa emission lines often show higher equivalent width (EW) in type 2 AGNs, presumably due to the attenuation of the underlying continuum \citep[e.g.][]{lamassa09}. 
Particularly, in type 2 AGNs with extreme X-ray obscuration (i.e. Compton-Thick), the narrow \FeKa lines even appear as the most prominent feature in the hard X-ray spectra and could have EW as large as several keV \citep[e.g.][]{levenson06}.
 As the narrow \FeKa emission likely arises from the putative torus rather than the BLR, studies of the narrow core could place constraints on the unified model.

Although the luminosity of the \FeKa line has been suggested to be an indicator of intrinsic AGN flux \citep{ptak03,levenson06,lamassa09}, it is still unclear whether the narrow \FeKa line emission itself (i.e. the line flux but not line EW) depends on inclination. To answer this question, we present in this paper a first systematic comparison of the narrow  \FeKa lines  in type 1 and type 2 AGNs. To perform such comparison, an intrinsic luminosity indicator of the central engine is required. Hard X-ray luminosity or reddening-corrected [O III] were commonly used as isotropic luminosity indicators in various studies \citep[e.g.][]{bassani99,diamond09,melendez08a}.
However, in the case of Compton-Thick  sources which we are interested in particularly, hard X-ray luminosity is also strongly attenuated due to Compton scattering. The dust reddening to [O III] brings extra uncertainty to the measurement of intrinsic [O III] luminosity. Furthermore, the [OIII] emission could be partially obscured in type 2 AGNs \citep{zhangkai08}, making reddening correction procedure more difficult.
In this paper, we select to use [OIV] 25.89 $\mu$m line as an intrinsic luminosity indicator. Since [OIV] 25.89 $\mu$m has relatively higher ionization potential (54.9 eV), it's less affected by star formation. It is also significantly less affected by extinction than [OIII] \citep[$A_v \sim 3 - 9$ corresponds to $A_{25.89\mu m} \sim 0.06 - 0.18$;][]{goulding09}.
\citet{melendez08a} have found that both [O IV] and [O III] luminosity correlate well with the very hard X-ray luminosity (14-195 keV) as measured by the SWIFT/BAT. Meanwhile \cite{diamond09} reported that the [OIV] luminosity distributions are indistinguishable for obscured and unobscured Seyferts, while [OIII] luminosities are systematically smaller for obscured Seyferts. These confirm that [OIV] could serve as a reddening-free isotropic luminosity indicator.
Also, [OIV] represents an improvement over the use of infrared continuum given the difficulty in isolating the AGN continuum from the host galaxy emission \citep{lutz04}. 

\section{Sample and Data Reduction}
\subsection{Sample Selection}
[OIV] 25.89 $\mu$m emission lines have been detected in large number of AGNs by the Infrared Spectrograph (IRS) on board the Spitzer Space Telescope in the first Long-Low (LL1, $19.5\sim38.0 \mu m$) IRS order. \citet{diamond09} and \citet{melendez08b} presented two large AGN samples (respectively 89 and 103 Seyferts) with [OIV] measurements. We cross-correlated the two samples with 2XMM catalogue (the second comprehensive catalogue of serendipitous X-ray sources from \xmm), which was released on 2007 August 22nd \citep{2xmm}. By excluding sources with less than 200 total EPIC counts in 2-12 keV, our final sample consists of 182 $XMM$ observations of 89 Seyferts. 
 We note that the final sample is never a homogeneous one, since it relies on the availability of both archival Spitzer spectral data and XMM data of individual sources. However, most (if not all) of the Spitzer and XMM observations were planned independent of their OIV emission and  \FeKa line emission. Our study, which is focus on OIV and  \FeKa line emission, is thus free from the diverse selection effects. The cutoff in X-ray counts ($>200$ counts, 2-12 keV)  excludes weaker X-ray sources ( compared with OIV emission). However, this effect is also independent  of  \FeKa line, since even for Compton-thick sources with  \FeKa EW of 1 keV, the emission line could only contribute 13\% to 2-12 keV EPIC counts (assuming a pure reflection model $pexrav$ with a photon index of 2 and a folded energy of 100 keV, plus a narrow gauss line at 6.4 keV whose EW is set to be 1 keV in xspec ).\\

We gathered the redshifts and coordinates for each source from Simbad. X-ray classification of type 2 AGNs (to distinguish between Compton-thin and Compton-thick) requires not only proper spectral fitting to full band X-ray spectra, but also much more complex models than the ones adopted in this paper, including partial covering absorber, soft X-ray excess, scattering component, contamination from host galaxy, etc. Furthermore, low quality spectra could make the situation more difficult, and in many cases one need additional diagnostics, such as  \FeKa line EW and T ratio ($F_{2-10 keV}/F_{[OIII]}$, see Bassani et al. 1999).  Thus a uniform  summary of X-ray classifications is unavailable.  Fortunately detailed X-ray studies for all but one Seyfert 2 galaxies in our sample have been published, and in this paper we choose to quote their X-ray classifications from the literature.  

The X-ray absorption column density of Seyfert 2 galaxy NGC 777 was not available from literature, and we classified it as Compton-Thin based on spectral fitting to XMM data. We divided the sample into 3 subsamples: Seyfert 1s (including Sy 1.2-1.5s,  33 sources), Compton-Thin Seyfert 2s (including Sy1.8-1.9s,  35 sources) and Compton-Thick Seyfert 2s (with $N_H > 10^{24} cm^{-2}$,  21 sources). 
In panel A of Fig. \ref{hist} we plot the histogram distribution of [OIV] luminosity for three subsamples.
Excluding 6 upper limits of $L_{[OIV]}$, we processed KS test (short for Kolmogorov-Smirnov test, which compares two samples under the null hypothesis that the samples are drawn from the same distribution \footnote{ A statistically significant difference corresponds to a confidence level of 95\% or higher}) on [OIV] luminosity between subsamples.
No significant difference was found (the differences between Sy1s and Compton-Thin Sy2s, Sy1s and Compton-Thick Sy2s, Compton-Thin Sy2s and Compton-Thick Sy2s are at confidence levels of respectively 89.2\%, 14.6\%, 85.5\%, with corresponding D values of 0.29, 0.17, 0.32), ensuring we are comparing subsamples in same intrinsic luminosity range.
Throughout this paper, we adopt cosmological parameters $H_0 = 70 km s^{-1} Mpc^{-1}$, $\Omega_m = 0.27$, and $\Omega_\lambda = 0.73$. 
The sample is given in Table \ref{data}.\\

\subsection{Data Reduction}
We used {\it XMM-SAS} (Science Analysis Software) version 9.0.0 and calibration files as of 2009 August to reduce archival data. Each observation was processed using the pipeline ``epchain" and ``emchain".
All the events were filtered to include only those with {\it XMM-SAS} quality flag {FLAG==0 (\#XMMEA\_EM)} for PN ( MOS1 and MOS2), and cleaned from flaring background with the task ``espfilt". The task ``espfilt" failed in 1 out of 9 exposures of Mrk 3, and this exposure (0009220301PNU002) was excluded from our further analysis.
Another 4 observations of 3C 273 in PrimeFull mode were eliminated due to serious pileup.
Generally we extracted event files comprising single- and double-pixel events (PATTERN $\leq$ 4) for PN, and single- to quadruple-pixel events (PATTERN $\leq$ 12) for  MOS1 and MOS2. However, in 90 exposures (Table \ref{data}), moderate pileup was identified using the task ``epatplot", therefore only single-pixel events were extracted for these 90 exposures to reduce the pileup effect \citep{ballet99}.

The level of pileup was also estimated through PIMMS\footnote{http://heasarc.nasa.gov/docs/software/tools/pimms.html} using observed spectra, and we found a maximum pileup fraction of 4\% in PN detectors in our sample.  Although such level of pileup could still alter observed spectral shape, the effect on the fitting to narrow \FeKa line, which is the aim of this work, is negligible (also see \S2.3).

We defined the source region as a circle with radius of 40\arcsec\ centered at the source position, and the background regions as three circles with the same radii around the source region, being central symmetric as much as possible, and kept away from the CCD edges, the out-of-time events strips and other sources (as illustrated in Fig. \ref{region}).
 When we were unable to avoid these disruptive features, the radii of the source or background regions were reduced down to 25\arcsec\ (see Table \ref{data}). We generated the source and background spectra together with the appropriate redistribution matrix and ancillary response file from the source and background regions for each exposure, using the task ``especget".
 Then for each source, to make the \FeKa feature prominent as it may not appear in individual observations, the PN spectra and the MOS1, MOS2 spectra from different observations were respectively combined, using the Ftools task ``addspec", (see Section 3.2 for further discussion on the combination). The combined MOS1 and MOS2 spectra will be referred to as MOS hereafter.

\subsection{ Spectral Fitting}
All spectra were grouped so that each channel contains at least one count, 
and spectral fitting were performed with C-statistic using Xspec version 
12.4.0. 
As our major goal in this work is to measure the flux of the narrow \FeKa line, 
we fitted the 5 -- 10 keV spectra with a simple model uniformly.
A Gaussian line was used to fit the narrow \FeKa line, and the continuum were 
fitted with an absorbed powerlaw ({\it zwabs*powerlaw+zgauss}).
Some examples of our spectral fitting are given in Fig. \ref{spectra}.

 Whenever available, PN and MOS spectra were fitted simultaneously with same model parameters, except for the powerlaw and Gaussian line normalizations, which were set free to account for the discrepancy among detector calibrations.  Normalizations from PN spectra are presented in these cases.  Pileup effects would be more significant in MOS detectors, which could alter the continuum spectral shape. For these MOS data suffering obvious pileup (see \S2.2), continuum parameters (photon index and absorption) were also setting free. 

In Table \ref{data} we list the best-fit parameter for the narrow \FeKa lines. For  25 of the sources, the detection of narrow \FeKa lines were insignificant (with  F-statistic confidence level $<99\%$). In  NGC7314 and I ZW 1, the centroid energies  (rest-frame) of the detected gauss line are larger than 6.4 keV at $>3\sigma$ level\footnote{$3\sigma$ errors of centroid energies were calculated through $xspec$, but only 90\% errors were presented in the tables.}, which seem to be due to ionized Fe lines, and extra gauss line at $\sim$ 6.4 keV is statistically not required. For these 27 sources  where the detection of the \FeKa line was not significant, we provide upper limit to their narrow \FeKa line flux by fitting with a gauss line with central energy fixed at 6.4 keV  (rest-frame) and the $\sigma$ at 44 eV (the  median $\sigma$ of the detected lines  for the full sample, see panel B of Fig. \ref{hist}). Fixing the energy and $\sigma$ to the median values found for each source's corresponding sub-population does not alter our results in this paper. In several sources, extra gaussian lines were statistically required (with F-statistic confidence $>99\%$), most of these lines could be attributed to either broader \FeKa line, ionized \FeKa line or Fe K$\beta$ line (see Table \ref{addgauss}). \\

The spectral fitting to the narrow \FeKa line could be significantly affected by continuum complexities. To quantify this effect, we  generated 390 artificial spectra based on various continuum models plus narrow lines, and apply our uniform model ({\it wabs*powerlaw+zgauss}) to fit the artificial spectra between 5 and 10 keV. The models we adopted for simulations include A) partially absorbed powerlaw ({\it pcfabs*powerlaw+zgauss} model in Xspec, with covering factor varying from 0 to 1, and $n_H$ between $10^{21}\sim10^{24} cm^{-2}$), and B) absorbed powerlaw plus reflection component ({\it wabs*powerlaw+pexrav+zgauss} in Xspec, with $n_H$ varying between $10^{21}\sim10^{24} cm^{-2}$, and relative strength of the direct component to the reflected component varying from 0 to 10).  The line width and EW of the simulated narrow \FeKa line were selected to cover the whole range of the  final best-fit parameters in our sample (see Table \ref{data}). Through the simulations we find that by restricting the fitting to 5 -- 10 keV range, we can recover the simulated narrow \FeKa lines with systematical deviation $<$ 2\%.  Adopting wider spectral range, such as 2 -- 10 keV, would yield significant bias to the narrow \FeKa line measurement (as large as 300\%), due to the improper fitting to the underlying continuum. 

\section{Statistical Distributions}

\subsection{Centroid Energy, Line Width and Line Equivalent Width}
\begin{figure}
\plotone{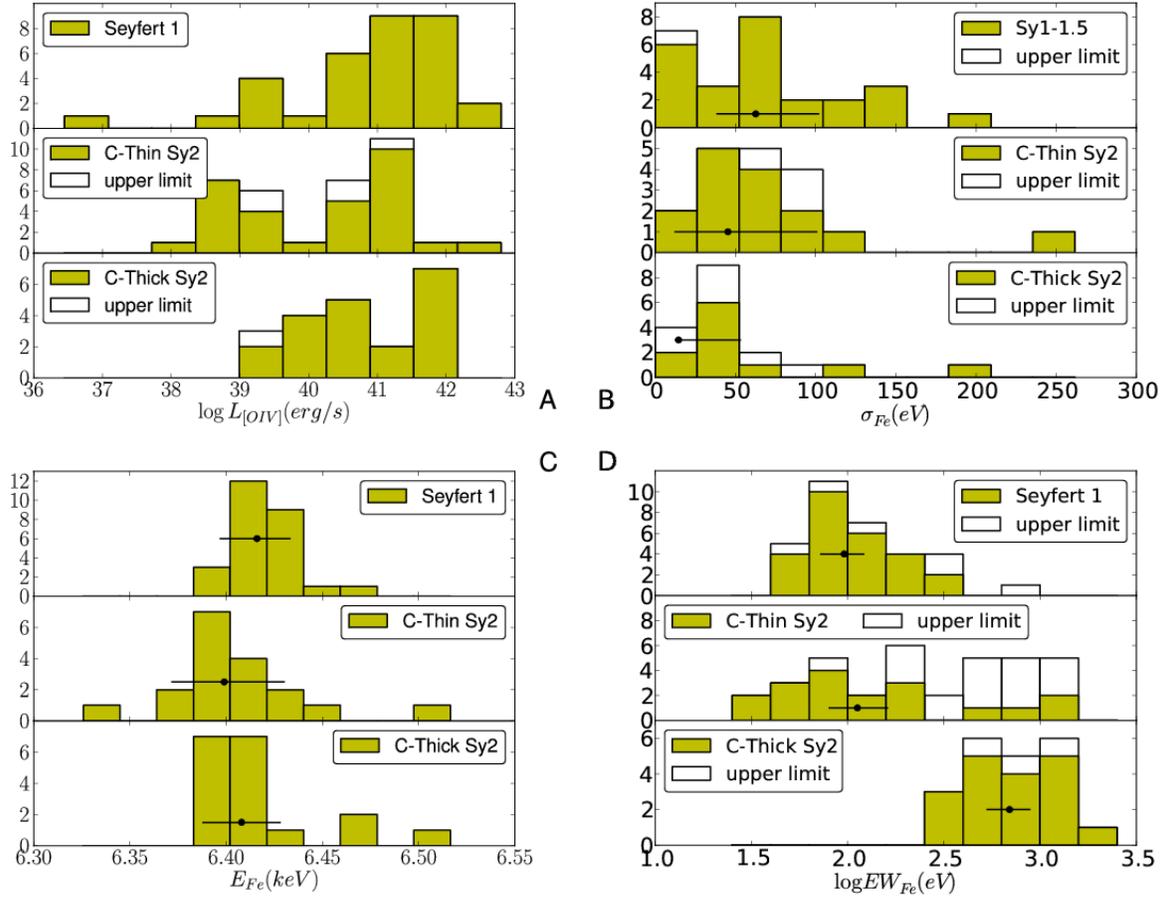}
\caption{\label{hist}The histograms of [OIV] line luminosities and centroid narrow \FeKa line energies, $\sigma$ and EW.
In panel B, C and D, we plot the  median values and errorbars of the sources with narrow \FeKa line detected.
}
\end{figure}

\begin{figure}
\plotone{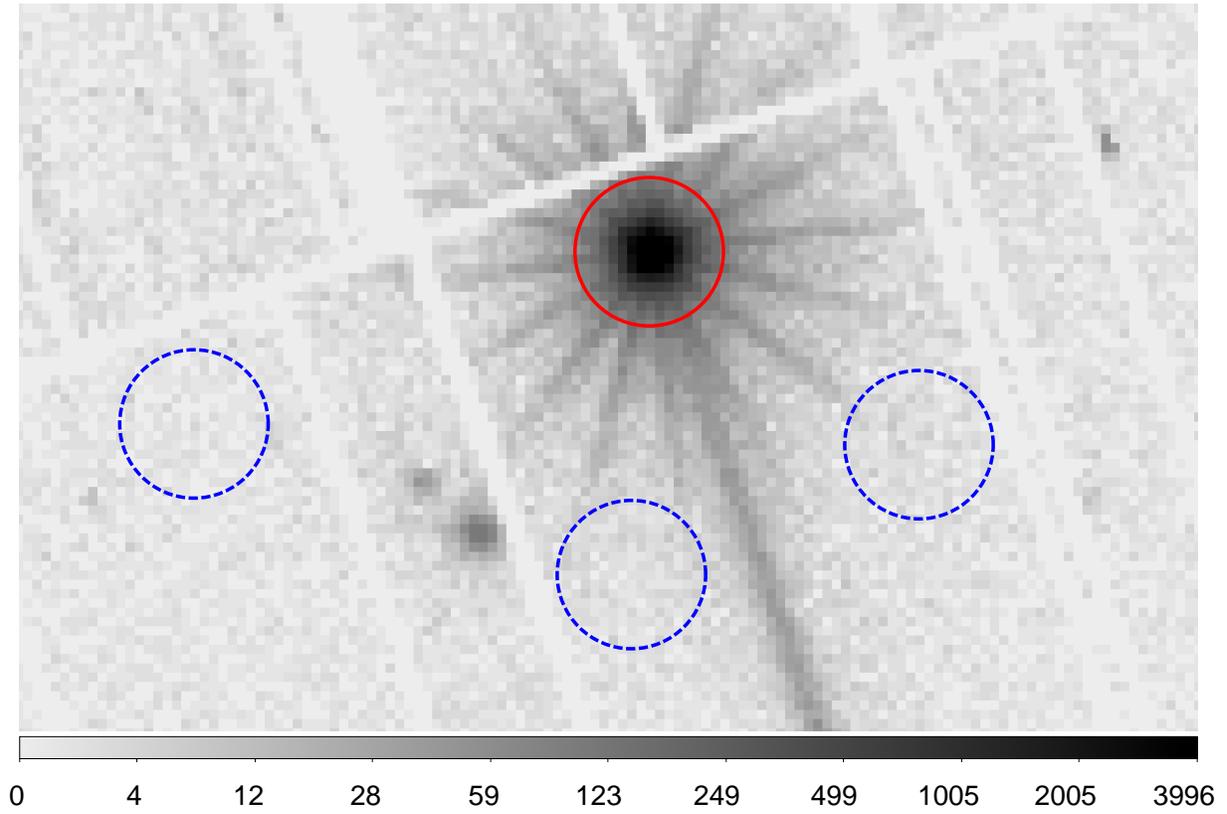}
\caption{\label{region}Example of source and background regions used to extract spectra (NGC 2992, ObsID 0147920301, PN detector),  with single-pixel events only. The colorbar refers to the photon counts in each pixel.}
\end{figure}

\begin{figure}
\plotone{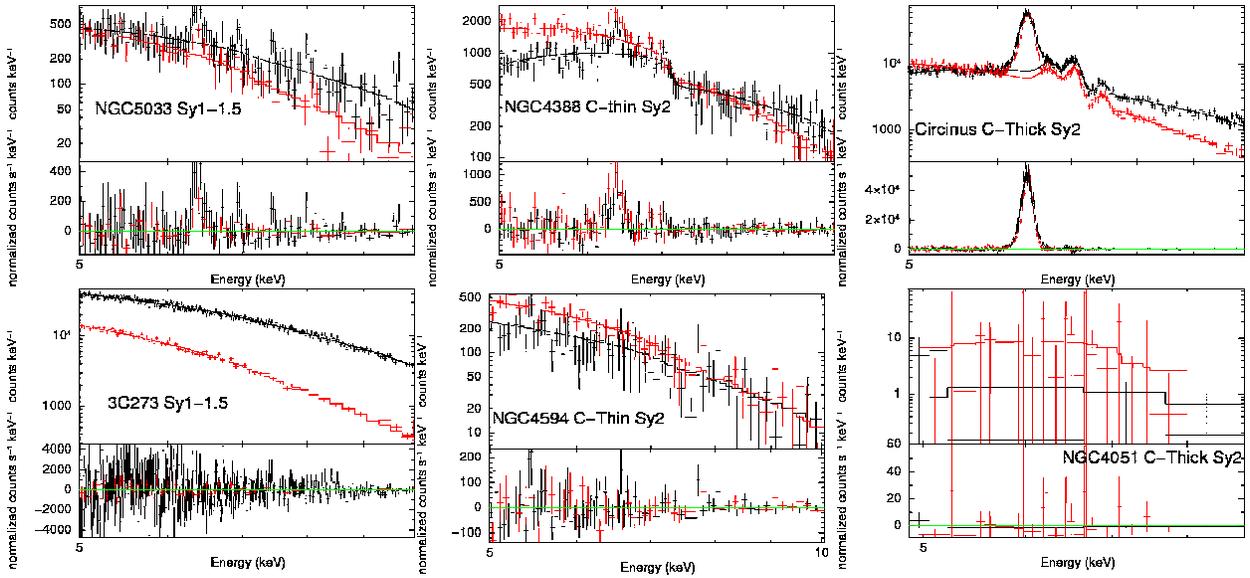}
\caption{\label{spectra}Sample spectra of our sample, including two Seyfert 
1-1.5 galaxies, two Compton-thin Seyfert 2 galaxies, and two Compton-thick 
Seyfert 2 galaxies.  The black and red data points refer to PN and MOS spectra respectively.
 All the model components but the narrow \FeKa line are over plotted to show the \FeKa line in the data to model ratio plot.
The upper three sources have \FeKa line detected, and the lower three do not.
}
\end{figure}

In panel C of Fig. \ref{hist} we first plot the centroid energy distribution for narrow \FeKa lines in three subsamples. We find while the line centroid energies in type 2 AGNs are generally consistent with 6.4 keV, those in type 1 AGNs tend to be slightly higher. KS test indicates that the difference in the narrow \FeKa line central energy distributions between type 1 AGNs and type 2 AGNs is statistically marginal (with a confidence level of $98.4\%$  and D=0.385). We also derived a weighted mean centroid energy of 6.409$\pm0.002$ keV for type 1 AGNs, and 6.402$\pm0.004$, 6.401$\pm0.002$ keV for Compton-Thin and Compton-Thick Sy2s respectively. Higher centroid energies  of the detected \FeKa lines of Sy1s listed in Table 1 could be due to contamination from  unresolved highly ionized Fe line or inner region (such as the accretion disk) component which is Doppler broadened and affected due to general relativity \citep{fabian89}\footnote{Modeling the broad component is beyond the scope of this paper. Systematic studies of broad \FeKa line emission from accretion disk are available in \citet{brenneman09} and \citet{nandra07}.}, and such components could be weaker in type 2 AGNs due to strong obscuration from the torus.

We note that by studying large samples of type 1 AGNs with HETG observations, Yaqoob \& Padmanabhan (2004) and Shu et al. (2010) have demonstrated that the centroid  \FeKa line energies in type 1 AGN strongly peak at 6.4 keV.  This indicates that the contamination to the narrow  \FeKa line core is much weaker in HETG spectra due to its much better spectral resolution.

We also compared the \FeKa line widths between type 1 and type 2 sources. Due to the limited spectral resolution of XMM PN/MOS, many of the narrow lines are unresolved, and only upper limits were given (see Table \ref{data}). The line width distributions are plotted in panel B of Fig. \ref{hist}, where can see slightly larger line width in type 1 AGNs  compared with type 2 sources.
Making use of ASURV (The Astronomy Survival Analysis; \citealp{feigelson85}), which can be used in the presence of upper limit data, and will be used in this paper whenever upper limits exist in the data, we performed Peto-Prentice Generalized Wilcoxon test, and found that the line width distribution in type 1 AGNs differs from that of type 2 AGNs with a confidence level of 92\%.
This pattern is consistent with the centroid energy distribution, suggesting the narrow \FeKa line in XMM spectra of type 1 AGNs is likely polluted by either highly ionized Fe line or broader emission from smaller radius.

Consistent with previous studies \citep[e.g.][]{bassani99}, we do find larger  \FeKa line equivalent width in type 2 AGNs (see panel D of Fig. \ref{hist}).
Through Peto-Prentice Generalized Wilcoxon test (ASURV), we find the  \FeKa line equivalent width distribution in type 1 AGNs differs from that of Compton-thin and Compton-thick Sy2s with confidence level of 70\% and $>$99.99\% respectively.

\begin{figure}
\epsscale{.80}
\plotone{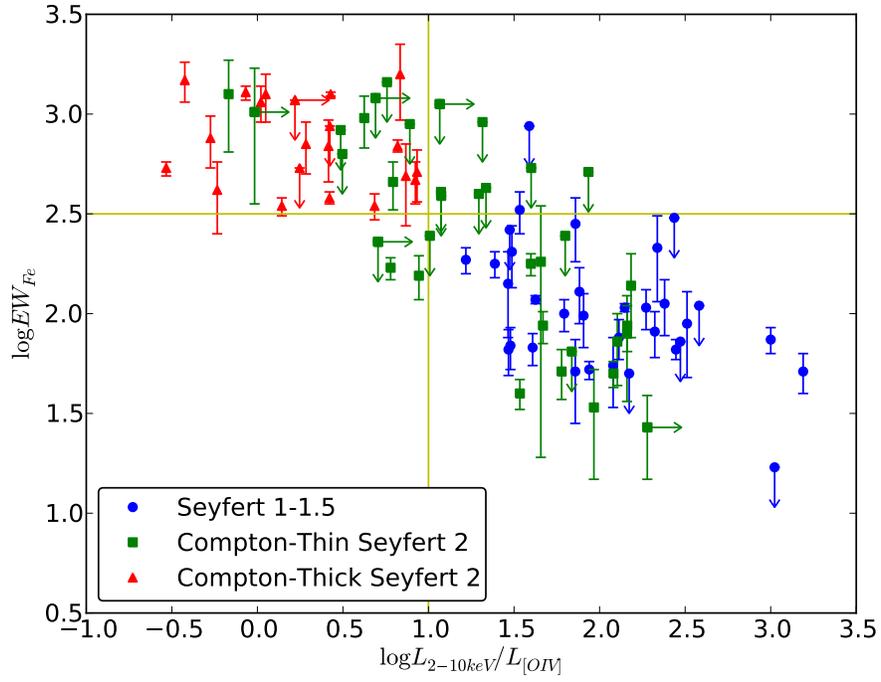}
\caption{\label{ewt}Narrow \FeKa line equivalent width versus $L_{2-10 keV}/L_{[OIV]}$.  The yellow lines can be used as plausible boundaries for Compton-Thick sources.}
\end{figure}

To examine the quoted X-ray classification, we reproduced the anti-correlation between the \FeKa equivalent width and the 2-10 keV luminosity normalized to the intrinsic AGN luminosity found by \citet{bassani99}, replacing reddening corrected [OIII] luminosity with [OIV] luminosity as proxy of the intrinsic luminosity (see Fig. \ref{ewt}). The 2-10 keV luminosities were calculated by fitting the 2-10 keV band spectra with two independent absorbed powerlaw and a gaussian line. We note such fitting could be unphysical to many sources, and is only valid to estimate the observed 2 -- 10 keV fluxes. In Fig. \ref{ewt}, the concentration of Compton-Thick sources at the upper left corner confirmed their classifications, including three Compton-Thick sources (NGC3982, NGC4501, NGC7674) with \FeKa line non-detected (upper limits to  \FeKa line EW were plotted).

\citet{bassani99} adopted a threshold of F$_{2-10keV}$/F$_{[OIII],corr}$ $<$ 1 to identify Compton-thick sources. \citet{lamassa10} reported an average ratio of Log(F$_{[OIV]}$/F$_{[OIII],corr}$) = -0.82 for 12$\micron$ selected Sy2s.
Thus $\log L_{2-10 keV}/L_{[OIV]} < 0.82$ 
can also be used as a boundary for Compton-Thick sources in our sample. 
In Fig. \ref{ewt} we plot $\log L_{2-10 keV}/L_{[OIV]} = 1$ and $\log EW_{Fe}=2.5$ (which include all Compton-thick sources) to delineate the Compton-thin/Compton-thick boundaries. Note there are four Compton-Thin sources fall in the Compton-Thick region.
They turn out to be "nearly" Compton-Thick with $\log n_H > 23$ (NGC1358: 23.4; NGC7479: 23.6; MCG-2-8-39: 23.7; MRK273: 23.8. See references in Table. \ref{data}).
We conclude that the quoted X-ray classifications are generally consistent with the measured \FeKa line EW and $\log L_{2-10 keV}/L_{[OIV]}$.
Re-classifying Seyfert 2 galaxies based on Fig. \ref{ewt} will not alter the results presented in this paper.

\subsection{Line Luminosity}

\begin{figure}
\epsscale{.90}
\plotone{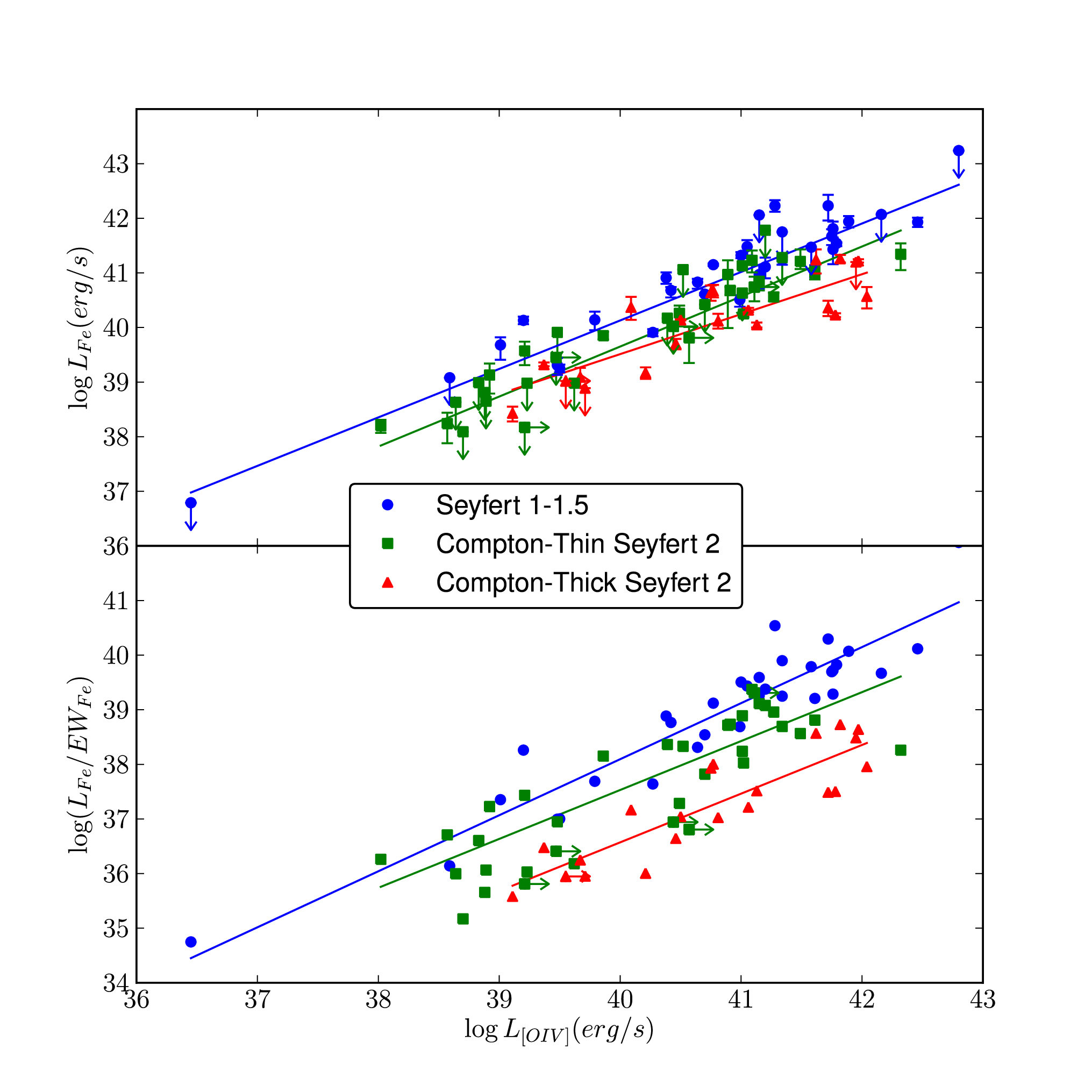}
\caption{Upper panel: Narrow \FeKa line luminosity versus $L_{[OIV]}$.
For the sources with narrow \FeKa lines detected, $90\%$ errors of the luminosities are plotted.
When the narrow \FeKa line is not detected, 3$\sigma$ upper limit of the luminosity is given.
Lower panel: 6.4 keV monochromatic luminosity versus $L_{[OIV]}$.
In both panels, the best-fit lines for Sy1s, Compton-Thin Sy2s and Compton-Thick Sy2s are shown as blue, green and red lines.
\label{o4_fe}}
\end{figure}

In the upper panel of Fig. \ref{o4_fe}  we plot the narrow \FeKa line luminosity versus [OIV] line luminosity for our samples. We first perform Buckley-James linear regression on the $\log L_{Fe} - \log L_{[OIV]}$ distribution for each subsample with ASURV. Six sources with only $L_{[OIV]}$ upper limit were shown in Fig. \ref{o4_fe}, but excluded from statistical analysis since ASURV can not handle upper limits in both data sets.
The best-fit slopes are:
\begin{tabbing}
\ \ \ \ \ \ \ \ \ \ \ \ \= Sy1-1.5:\ \ \ \ \ \ \ \ \ \= $\log L_{Fe} = \log L_{[OIV]}\times (0.89\pm 0.09)+4.62\pm 0.40$\\
\>	C-Thin Sy2:	\>$\log L_{Fe} = \log L_{[OIV]}\times(0.92\pm0.08)+3.02\pm0.36$\\
\>	C-Thick Sy2:	\>$\log L_{Fe} = \log L_{[OIV]}\times(0.73\pm0.12)+10.24\pm0.45$\\
\end{tabbing}
We note that excluding the extremely faint Seyfert 1 M81 does not significantly alter the best-fit slope and other results presented in this paper.\\ 

To explore the difference in $L_{[Fe]}$  between subsamples, we compare the residuals of $\log L_{Fe}$ from the best-fit line of $\log L_{Fe}-\log L_{[OIV]}$ for Sy1s for each subsample. This is valid since the three subsamples show consistent $L_{[OIV]}$ distributions (see \S2.1 and Fig. \ref{diff}). With Peto-Prentice Generalized Wilcoxon Test (ASURV), we found significant differences between Sy1s and Compton-thin Sy2s, and between Sy1s and Compton-thick Sy2s in the $\log L_{Fe}$ residuals (with confidence level of $99.98\%$ and $>99.99\%$ respectively, see Fig. \ref{diff}). This implies the distribution of $\log L_{Fe}-\log L_{[OIV]}$ also differs statistically significantly between Sy1s and Sy2s. The difference in $\log L_{Fe}$ residuals between Compton-Thin and Compton-Thick Sy2s is only marginal (with a significance level of $ 86\%$). 
By comparing the mean values of the $\log L_{Fe}$ residuals in subsamples we find the narrow \FeKa lines in Sy1s are $2.9^{+0.8}_{-0.6}$ times stronger  in terms of luminosity than that of Compton-Thin Sy2s, and $5.6^{+1.9}_{-1.4}$ times stronger than that of  Compton-Thick Sy2s. 
Simply comparing $\log L_{Fe} / \log L_{[OIV]}$ between subsamples yields consistent results, that the narrow \FeKa lines in Sy1s are $2.6^{+0.8}_{-0.6}$ times stronger  than that of Compton-Thin Sy2s, and $5.7^{+2.1}_{-1.6}$ times stronger than that of Compton-Thick Sy2s.

For comparison, we plot 6.4 keV monochromatic luminosity ($L_{Fe}/EW_{Fe}$) versus $L_{[OIV]}$ for three subsamples in the lower panel of Fig. \ref{o4_fe}.  Adopting the same methods used above, we find that the best fitting slopes for Sy1s, Compton-Thin Sy2s and Compton-Thick Sy2s are respectively 1.03$\pm$0.07, 0.90$\pm$0.10 and 0.89$\pm$0.11, 
and the 6.4 keV monochromatic luminosity in Sy1s are $4.2^{+1.6}_{-1.2}$ times stronger than that of Compton-Thin Sy2s, and $43.2^{+16.0}_{-11.7}$ times that of Compton-Thick Sy2s.  These decrements, if simply attributed to photo-electric absorption, require column densities of $8.6^{+1.9}_{-2.1}\times10^{23} cm^{-2}$ and $2.3^{+0.2}_{-0.2}\times10^{24} cm^{-2}$ respectively (calculated using a simple photo-electric absorbing model {\it wabs*gauss} in Xspec, with the line energy and $\sigma$ fixed at 6.4 keV and 0 eV).

To examine whether our fitting could be biased by spectral combination which could produce spurious results for variable sources, we repeat our spectral fit to individual observations. Discarding the observations in which narrow \FeKa line is no longer detectable, we calculated the weighted mean line flux for each source. Fig. \ref{fefluxes} compares the weighted mean line flux from individual observations with the line flux from the combined spectra, between which no obvious difference is found.

\section{Discussion}
By normalizing to [OIV], which serves as an isotropic indicator of AGN luminosity, we find that the narrow \FeKa line in Compton-Thin and Compton-Thick Sy2s  are  $2.9^{+0.8}_{-0.6}$ and $5.6^{+1.9}_{-1.4}$  times weaker than Sy1s respectively. 
We note that based on the X-ray reprocessing torus model of \citet{krolik94}, \citet{levenson06} performed numerical simulations to 7 Compton-Thick AGNs to estimate their intrinsic 2 - 10 keV luminosity from observed \FeKa EW and luminosity. In their model, the obscuring torus is cylindrical symmetric, filled with constant density material, and has a square or rectangular cross-section allowing for unobstructed views over the half-opening angle. They obtained a typical value of $L_{Fe}/L_{intrinsic(2-10  keV})$ = $2\times10^{-3}$ for the 7 Compton-Thick AGNs.
Such ratio is approximately 5 times smaller than the median value of our Sy1s with \FeKa line detected (a median line EW of 96 eV could be converted to $F_{Fe}/F_{2-10  keV}$ = $9.7\times10^{-3}$ assuming a powerlaw with photon index $\Gamma=1.9$).  Taking our observed $L_(2-10  keV)$ for Sy1s presented in Table 1, we obtained a similar $F_{Fe}/F_{2-10  keV}$ = $11.2\times10^{-3}$, 5.6 times larger than those for Compton-thick Sy2s presented  by Levenson et al. These are remarkably consistent with our results that the observed \FeKa emission in Compton-Thick AGNs is $5.6^{+1.9}_{-1.4}$ times weaker than type 1 AGNs.

\begin{figure}
\epsscale{.90}
\plotone{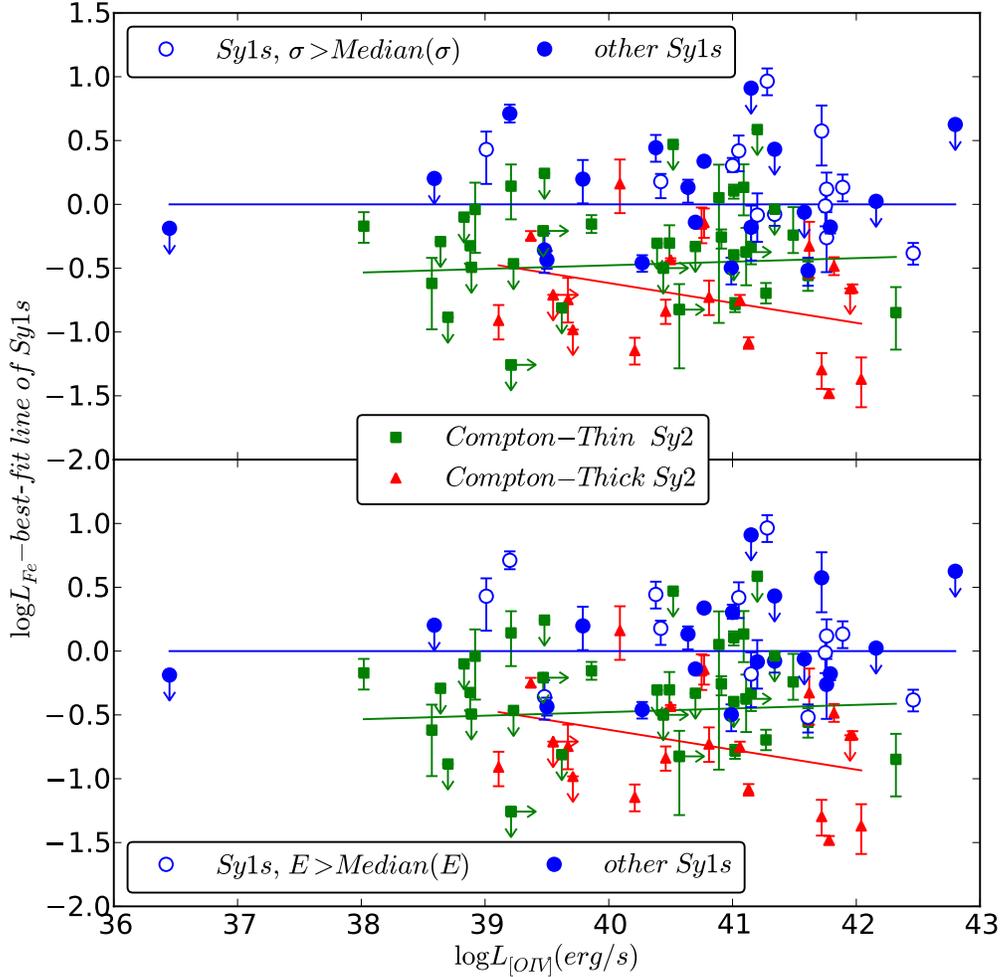}
\caption{\label{diff}
The deviation of $\log L_{Fe}$ from the best-fit $\log L_{Fe}$--$\log L_{O[IV]}$ line of Seyfert 1 galaxies, where we clearly see weaker narrow \FeKa emission line in Seyfert 2 galaxies. Best-fit $\log L_{Fe}$--$\log L_{O[IV]}$ correlations for Compton-thin and Compton-thick Sy2s are over-plotted.
We also divide Sy1s by the \FeKa line width and centroid energy using the median value of the  Sy1s with \FeKa line detected as a boundary, which demonstrate that Sy1s with larger \FeKa line width or centroid energy do not show obviously stronger line emission.
}
\end{figure}

\begin{figure}
\epsscale{.90}
\plotone{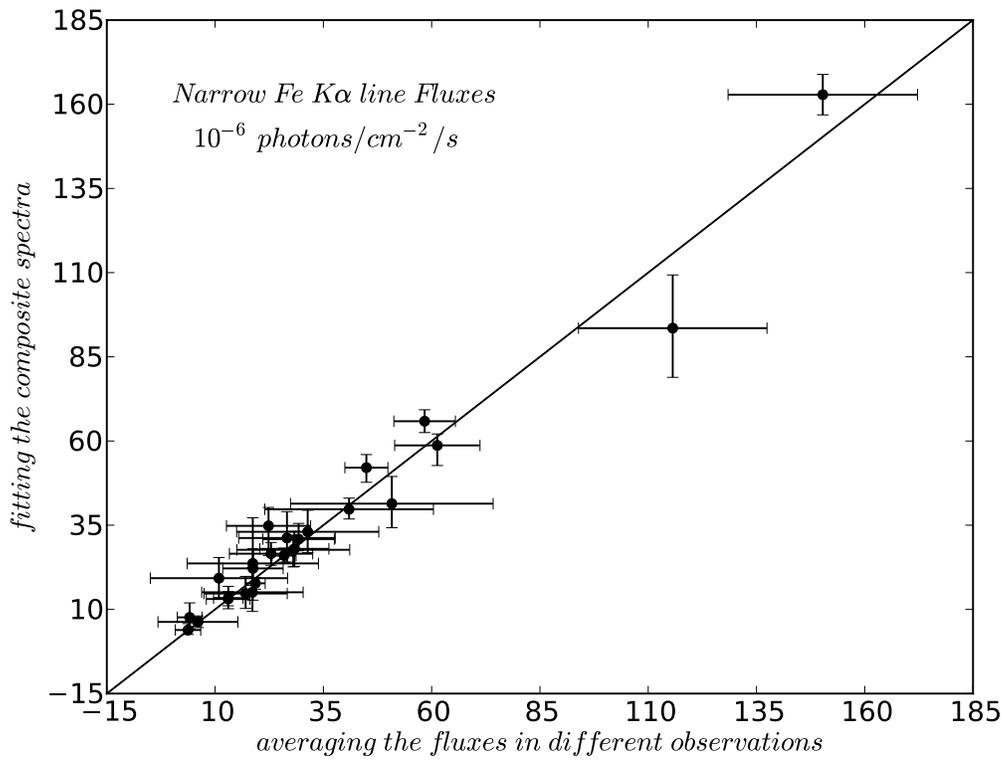}
\caption{\label{fefluxes}For sources with multiple XMM observations, 
spectral fitting to composite spectra yield narrow \FeKa line
flux consistent with the weighted mean of individual exposures.
 The line represents a slope of unity.
}
\end{figure}

Meanwhile we also find marginally higher centroid energy and larger line width in type 1 AGNs  compared with type 2 AGNs. This indicates the narrow \FeKa line in XMM spectra of type 1 AGNs is likely polluted by highly ionized Fe line or broader component from smaller radius in type 1 AGNs, and such contamination is weaker in type 2 AGNs due to the torus obscuration.
If highly ionized Fe line or broader component from smaller radius was responsible for the disparity in \FeKa luminosity between Sy1s and Sy2s, the Sy1s with larger line widths and/or higher centroid energy should have systematically higher $L_{Fe}/L_{[OIV]}$ ratios.
In Fig. \ref{diff} we see that Sy1 galaxies with higher  \FeKa line centroid energy or larger line width ( compared with the median values of the Sy1s with \FeKa line detected) do not have obviously larger $L_{Fe}$/$L_{[OIV]}$.
KS tests show that the differences in $\log L_{Fe}$ residuals in Fig. \ref{diff} between Sy1s with higher and lower centroid energy, and between Sy1s with larger and smaller line widths, are at confidence levels of 51\% (D=0.31) and 91\% (D=0.46) respectively, neither of which are statistically significant. 
Similar confidence levels of 51\% (D=0.31) and 77\% (D=0.38) were obtained if we simply compare Log($L_{Fe}$/$L_{[OIV]}$) of Sy1 subsamples. The average Log($L_{Fe}$/$L_{[OIV]}$) is $0.18\pm0.14$ for Sy1s with higher centroid energy, $-0.02\pm0.09$ for lower centroid energy, $0.16\pm0.11$ for Sy1s with larger line width, and $0.00\pm0.12$ for smaller line width.
This indicates that the contribution from either highly ionized Fe line or broader emission from much smaller radius is too small to explain the difference we found between type 1 and type 2 AGNs.

 We note that among the 26 Sy1s with narrow \FeKa line detected in table 1, eight have both higher line centroid energy (E$_c$) and larger line width ($\sigma$), and eight show lower E$_c$ and smaller $\sigma$ ( compared with the median values of Sy1 sample). Five Sy1s show higher E$_c$ but smaller $\sigma$, and the rest five have lower E$_c$ but larger $\sigma$.  Thus no significant correlation is found between E$_c$ and line width.  Meanwhile, contamination from unresolved ionized  \FeKa emission could produce lines with higher E$_c$ and larger $\sigma$, those lines with smaller E$_c$ and larger $\sigma$ are more likely due to Doppler broadening.  
 
The \FeKa line emission could be determined by many factors, including the geometry, column density, covering factor of the line-emitting gas, element abundances, and the orientation of the observer's line of sight relative to the line-emitting structure. In the standard unified model, the only difference between type 1 and type 2 AGNs is the orientation, thus is the only plausible factor to explain the different \FeKa line emission. Observational selection effects could possibly favor smaller column density and covering factor of the torus in type 1 sources, but both effects predict weaker \FeKa emission (contrarily, larger covering factor and column density of the torus could generally enhance the narrow \FeKa line emission in type 1 sources). Meanwhile, we do not expect significantly higher Fe abundance in type 1 AGNs, not to mention that the \FeKa line emission is insensitive to Fe abundance unless in optically-thin limit (N$_H$ $<$ 10$^{23}$ cm$^{-2}$, see \citealp{yaqoob09}). 

\citet{yaqoob09} presented Monte-Carlo simulations on the production of \FeKa emission in toroidal neutral X-ray reprocessor (also see \citealp{murphy09}). They clearly demonstrate that the strength of the line emission from a toroidal torus could dramatically depend on the viewing angle, as long as the column density of the reprocessor is significantly larger than 10$^{23}$ cm$^{-2}$.  For instance, for column density of 10$^{24}$ -- 10$^{25}$ cm$^{-2}$, \FeKa emission at face-on inclination could be around 3 -- 30 times stronger than edge-on inclination (see Fig. 2 of \citealp{yaqoob09}). This is mainly because of the larger optical depth to \FeKa emission due to photo-electric absorption  and Compton scattering at edge-on inclination. This scheme could easily produce the difference in the observed \FeKa line emission between type 1 and type 2 AGNs. Clearly, this scheme requires the existence of Compton-thick torus in both type 1 and type 2 AGNs, although the line of sight is free of Compton-thick obscuration in type 1 and Compton-thin type 2 AGNs.

However, if the line-emitting gas is clumpy instead of smoothly distributed, the inclination dependency of the line emission could be reduced or even smeared out (also see the \citealp{nandra94} and \citealp{miller09}), dependent  on the filling factor and covering factor of the clumpy blobs.  Monte-Carlo simulations of the dust emission and radiative transfer in the clumpy torus have been performed to model infrared emission in various types of AGNs, which could produce strong constraints on the clumpiness of the dusty torus \citep[e.g.][]{nenkova08}.  Our results could thus provide independent constraints to the distribution of the cold gas (which produces \FeKa line emission, and its distribution may differ from that of dust) in a clumpy torus model once detailed Monte-Carlo simulation of \FeKa line production in such model would be available.
 
Our results indicate that the narrow \FeKa emission in AGNs is anisotropic, with weaker emission (although higher EW) in obscured AGNs. Extra correction factors should be applied if one uses observed \FeKa emission to estimate AGN's intrinsic luminosity.

\acknowledgments

The work was supported by Chinese NSF through NSFC10773010/10825312, and the Knowledge Innovation Program of CAS (Grant No. KJCX2-YW-T05).

\bibliography{ms}

\begin{deluxetable}{lcccrrrlllrr}
\tabletypesize{\scriptsize}
\tablecaption{\label{data}Observations and Data}
\rotate
\tablewidth{0pt}
\tablehead{
\colhead{Name} &\colhead{z} &\colhead{ObsID} &\colhead{Pileup;Radii} &\colhead{$\log L_{2-10 keV}$} &\colhead{$\log L_{[OIV]}$} &\colhead{$\log L_{Fe}$} &\colhead{$E_{Fe} (keV)$} &\colhead{$\sigma_{Fe} (eV)$} &\colhead{EW (eV)} &\colhead{Type} &\colhead{Ref}
}
\startdata
3C 120         &0.033010 &0109131101 &;                 &44.07 &42.46  &$41.93_{-0.09}^{+0.08}$ &$6.42_{-0.02}^{+0.02}$  &$109.6_{-25.1}^{+22.6}$ &$67_{-12}^{+12}$        &S1 &\\
               &         &0152840101 &;                &       &                        &                        &                        &                        &   &\\
3C 273         &0.158339 &0112770101 &;                 &45.82 &42.80  &$<$43.24                  &6.40*               &44*                 &$<$17                     &S1 &\\
               &         &0112770201 &m2;              &       &                        &                        &                        &                        &   &\\
               &         &0112770501 &;                &       &                        &                        &                        &                        &   &\\
               &         &0112770601 &;                &       &                        &                        &                        &                        &   &\\
               &         &0112770701 &;                &       &                        &                        &                        &                        &   &\\
               &         &0112770801 &;                &       &                        &                        &                        &                        &   &\\
               &         &0112771001 &m2;              &       &                        &                        &                        &                        &   &\\
               &         &0112771101 &m2;              &       &                        &                        &                        &                        &   &\\
               &         &0126700101 &m1,m2,pn;        &       &                        &                        &                        &                        &   &\\
               &         &0126700201 &m1,m2,pn;        &       &                        &                        &                        &                        &   &\\
               &         &0126700301 &m1,m2;           &       &                        &                        &                        &                        &   &\\
               &         &0126700401 &m1,m2,pn;        &       &                        &                        &                        &                        &   &\\
               &         &0126700501 &m1,m2,pn;        &       &                        &                        &                        &                        &   &\\
               &         &0126700601 &m1,m2;           &       &                        &                        &                        &                        &   &\\
               &         &0126700701 &m1,m2;           &       &                        &                        &                        &                        &   &\\
               &         &0126700801 &m1,m2;           &       &                        &                        &                        &                        &   &\\
               &         &0136550101 &m1,m2;           &       &                        &                        &                        &                        &   &\\
               &         &0136550501 &m1,m2;           &       &                        &                        &                        &                        &   &\\
               &         &0136550801 &m1,m2;           &       &                        &                        &                        &                        &   &\\
               &         &0136551001 &m1,m2;           &       &                        &                        &                        &                        &   &\\
               &         &0159960101 &m1,m2;           &       &                        &                        &                        &                        &   &\\
               &         &0414190101 &m1,m2;           &       &                        &                        &                        &                        &   &\\
3C 390.3       &0.056100 &0203720201 &;                 &44.47 &41.28  &$42.23_{-0.11}^{+0.10}$ &$6.43_{-0.02}^{+0.02}$  &$96.2_{-34.0}^{+40.7}$  &$51_{-11}^{+13}$        &S1 &\\
               &         &0203720301 &;                &       &                        &                        &                        &                        &   &\\
Circinus       &0.001448 &0111240101 &;                 &40.93&40.50  &$40.14_{-0.01}^{+0.01}$ &$6.39_{-0.00}^{+0.00}$  &$44.1_{-3.7}^{+2.5}$    &$1251_{-22}^{+22}$      &CT &2\\
ESO 103-G035   &0.013286 &0109130601 &;                 &43.01&41.15  &$40.97_{-0.26}^{+0.17}$ &$6.47_{-0.03}^{+0.03}$  &$0.4_{-0.4}^{+53.9}$    &$51_{-23}^{+24}$        &S1 &\\
ESO 141-G055   &0.036000 &0101040501 &m1,m2;            &43.81&41.34  &$<$41.75                  &6.40*               &44*                 &$<$73                     &S1 &\\
IC 2560        &0.009757 &0203890101 &;                 &40.99&41.06  &$40.32_{-0.04}^{+0.04}$ &$6.41_{-0.01}^{+0.00}$  &$<3.6     $     &$1280_{-107}^{+114}$    &CT &11\\
IC 4329A       &0.016054 &0101040401 &m1,m2,pn;         &43.73 &41.79  &$41.54_{-0.05}^{+0.02}$ &$6.41_{-0.01}^{+0.01}$  &$52.8_{-12.6}^{+9.4}$   &$52_{-5}^{+3}$          &S1 &\\
               &         &0147440101 &m1,m2;           &       &                        &                        &                        &                        &   &\\
IRAS01475-0740 &0.017666 &0200431101 &;                 &41.77&40.70  &$<$40.42                  &6.40*               &44*                 &$<$405                    &S2 &25\\
IRAS15091-2107 &0.044607 &0300240201 &;                 &43.63&42.16  &$<$42.07                  &6.40*               &44*                 &$<$263                    &S1 &\\
I ZW 1         &0.061142 &0110890301 &;                 &43.75 &41.58  &$<$41.47                  &6.40*               &44*                 &$<$50                     &S1 &\\
               &         &0300470101 &;                &       &                        &                        &                        &                        &   &\\
M81            &-0.00011 &0112521001 &;                 &39.03 &36.45  &$<$36.79                  &6.40*               &44*                 &$<$110                    &S1 &\\
               &         &0112521101 &;                &       &                        &                        &                        &                        &   &\\
               &         &0200980101 &;M1:30           &       &                        &                        &                        &                        &   &\\
MCG-2-58-22    &0.046860 &0109130701 &;                 &44.23&41.72  &$42.23_{-0.27}^{+0.20}$ &$6.41_{-0.08}^{+0.08}$  &$145.0_{-74.8}^{+136.3}$ &$90_{-41}^{+53}$        &S1 &\\
MCG-2-8-39     &0.029894 &0301150201 &;                 &42.28&41.49  &$41.21_{-0.14}^{+0.22}$ &$6.41_{-0.06}^{+0.08}$  &$76.5_{-75.9}^{+81.1}$  &$456_{-123}^{+303}$     &S2 &26\\
MCG-5-13-17    &0.012445 &0312190701 &;                 &42.17&40.64  &$40.83_{-0.12}^{+0.06}$ &$6.41_{-0.02}^{+0.02}$  &$21.2_{-23.0}^{+30.8}$  &$334_{-80}^{+49}$       &S1 &\\
MCG-6-30-15    &0.007749 &0111570101 &m2;               &42.74 &40.42  &$40.68_{-0.13}^{+0.06}$ &$6.44_{-0.01}^{+0.01}$  &$141.5_{-10.0}^{+10.3}$ &$82_{-20}^{+12}$        &S1 &\\
               &         &0111570201 &m2;              &       &                        &                        &                        &                        &   &\\
               &         &0029740101 &m1;              &       &                        &                        &                        &                        &   &\\
               &         &0029740701 &m1;              &       &                        &                        &                        &                        &   &\\
               &         &0029740801 &m1;              &       &                        &                        &                        &                        &   &\\
MRK 3          &0.013509 &0111220201 &;                 &42.39 &41.97  &$41.22_{-0.03}^{+0.03}$ &$6.42_{-0.01}^{+0.01}$  &$38.7_{-10.7}^{+9.2}$   &$383_{-27}^{+31}$       &CT &2\\
               &         &0009220301 &;                &       &                        &                        &                        &                        &   &\\
               &         &0009220401 &;                &       &                        &                        &                        &                        &   &\\
               &         &0009220501 &;                &       &                        &                        &                        &                        &   &\\
               &         &0009220601 &;                &       &                        &                        &                        &                        &   &\\
               &         &0009220701 &;                &       &                        &                        &                        &                        &   &\\
               &         &0009220901 &;                &       &                        &                        &                        &                        &   &\\
               &         &0009221001 &;                &       &                        &                        &                        &                        &   &\\
               &         &0009221601 &;                &       &                        &                        &                        &                        &   &\\
MRK 6          &0.018813 &0144230101 &;                 &43.09 &41.61  &$41.04_{-0.12}^{+0.10}$ &$6.43_{-0.02}^{+0.02}$  &$0.6_{-0.6}^{+43.7}$    &$69_{-16}^{+17}$        &S1 &\\
               &         &0305600501 &;                &       &                        &                        &                        &                        &   &\\
MRK 79         &0.022189 &0103860801 &;                 &43.23 &41.76  &$41.43_{-0.27}^{+0.20}$ &$6.39_{-0.04}^{+0.06}$  &$70.4_{-71.6}^{+86.6}$  &$142_{-65}^{+81}$       &S1 &\\
               &         &0103862101 &;                &       &                        &                        &                        &                        &   &\\
MRK 231        &0.042170 &0081340201 &;                 &42.49&41.62  &$41.24_{-0.25}^{+0.19}$ &$6.46_{-0.11}^{+0.11}$  &$187.3_{-85.5}^{+133.4}$ &$486_{-212}^{+264}$     &CT &27\\
MRK 273        &0.037780 &0101640401 &;                 &42.15&42.32  &$41.34_{-0.29}^{+0.20}$ &$6.52_{-0.11}^{+0.09}$  &$261.6_{-123.8}^{+165.3}$ &$1249_{-615}^{+741}$    &S2 &28\\
MRK 335        &0.025785 &0101040101 &m1,m2,pn;         &43.43 &41.05  &$41.48_{-0.16}^{+0.12}$ &$6.44_{-0.03}^{+0.03}$  &$208.8_{-51.8}^{+56.4}$ &$113_{-34}^{+35}$       &S1 &\\
               &         &0306870101 &m1;              &       &                        &                        &                        &                        &   &\\
MRK 348        &0.015034 &0067540201 &;                 &43.19&41.09  &$41.23_{-0.22}^{+0.18}$ &$6.40_{-0.05}^{+0.03}$  &$57.1_{-56.5}^{+81.2}$  &$73_{-28}^{+38}$        &S2 &18\\
MRK 509        &0.034397 &0130720101 &;                 &44.00 &41.89  &$41.94_{-0.11}^{+0.10}$ &$6.42_{-0.03}^{+0.03}$  &$124.2_{-40.6}^{+48.5}$ &$76_{-17}^{+19}$        &S1 &\\
               &         &0130720201 &m1;M1M2:30,PN:35 &       &                        &                        &                        &                        &   &\\
MRK 573        &0.017179 &0200430701 &;                 &41.45&41.72  &$40.36_{-0.15}^{+0.13}$ &$6.39_{-0.04}^{+0.02}$  &$<42.1     $    &$754_{-216}^{+256}$     &CT &12\\
MRK 609        &0.034488 &0103861001 &;                 &42.63&41.34  &$<$41.28                  &6.40*               &44*                 &$<$398                    &S2 &19\\
MRK 841        &0.036422 &0112910201 &;                 &43.66 &41.75  &$41.67_{-0.16}^{+0.13}$ &$6.42_{-0.04}^{+0.04}$  &$131.4_{-46.1}^{+58.8}$ &$97_{-29}^{+33}$        &S1 &\\
               &         &0205340401 &;                &       &                        &                        &                        &                        &   &\\
               &         &0070740101 &;                &       &                        &                        &                        &                        &   &\\
               &         &0070740301 &;                &       &                        &                        &                        &                        &   &\\
NGC 424        &0.011764 &0002942301 &;                 &41.68&40.75  &$40.64_{-0.15}^{+0.13}$ &$6.39_{-0.02}^{+0.03}$  &$<38.7     $    &$514_{-148}^{+178}$     &CT &31\\
NGC 513        &0.019544 &0301150401 &;                 &42.55&40.89  &$40.97_{-0.98}^{+0.26}$ &$6.33_{-0.08}^{+0.08}$  &$118.1_{-118.7}^{+90.4}$ &$183_{-163}^{+153}$     &S2 &26\\
NGC 526A       &0.019097 &0109130201 &;                 &43.28 &41.20  &$41.11_{-0.21}^{+0.17}$ &$6.41_{-0.02}^{+0.03}$  &$75.1_{-75.5}^{+55.8}$  &$55_{-20}^{+26}$        &S1 &\\
               &         &0150940101 &;                &       &                        &                        &                        &                        &   &\\
NGC 777        &0.016728 &0203610301 &;                 &41.13 &$<$40.44 &$<$40.02                  &6.40*               &44*                 &$<$1213                   &S2 &1\\
               &         &0304160301 &;                &       &                        &                        &                        &                        &   &\\
NGC 985        &0.043143 &0150470601 &;                 &43.64&41.76  &$41.81_{-0.16}^{+0.13}$ &$6.44_{-0.03}^{+0.03}$  &$89.3_{-37.9}^{+37.6}$  &$128_{-40}^{+43}$       &S1 &\\
NGC 1068       &0.003793 &0111200101 &m1,pn;            &41.25 &41.78  &$40.23_{-0.04}^{+0.03}$ &$6.41_{-0.01}^{+0.00}$  &$33.0_{-9.5}^{+7.8}$    &$532_{-44}^{+39}$       &CT &2\\
               &         &0111200201 &m1,pn;           &       &                        &                        &                        &                        &   &\\
NGC 1194       &0.013596 &0307000701 &;                 &41.69&40.77  &$40.67_{-0.12}^{+0.11}$ &$6.40_{-0.02}^{+0.02}$  &$43.4_{-43.0}^{+38.3}$  &$468_{-114}^{+135}$     &CT &29\\
NGC 1275       &0.017559 &0305780101 &m1,m2,pn;         &43.39 &$<$41.11 &$40.74_{-0.26}^{+0.19}$ &$6.40_{-0.02}^{+0.03}$  &$<60.1     $    &$27_{-12}^{+15}$        &S2 &3\\
               &         &0085110101 &m1,m2;           &       &                        &                        &                        &                        &   &\\
NGC 1358       &0.013436 &0301650201 &;PN:25            &41.11&40.49  &$40.26_{-0.15}^{+0.14}$ &$6.44_{-0.03}^{+0.03}$  &$36.4_{-36.7}^{+50.4}$  &$956_{-278}^{+366}$     &S2 &5\\
NGC 1365       &0.005457 &0151370101 &;PN:25            &41.80 &41.02  &$40.25_{-0.06}^{+0.05}$ &$6.38_{-0.01}^{+0.01}$  &$91.2_{-11.2}^{+8.5}$   &$168_{-21}^{+21}$       &S2 &4\\
               &         &0151370201 &;                &       &                        &                        &                        &                        &   &\\
               &         &0151370701 &;                &       &                        &                        &                        &                        &   &\\
               &         &0205590301 &;                &       &                        &                        &                        &                        &   &\\
               &         &0205590401 &;                &       &                        &                        &                        &                        &   &\\
NGC 1386       &0.002895 &0140950201 &;                 &39.79&40.21  &$39.17_{-0.11}^{+0.10}$ &$6.39_{-0.02}^{+0.02}$  &$53.8_{-42.3}^{+29.7}$  &$1463_{-334}^{+383}$    &CT &6\\
NGC 2273       &0.006138 &0140951001 &;                 &40.92&40.09  &$40.37_{-0.23}^{+0.19}$ &$6.40_{-0.04}^{+0.04}$  &$<96.1     $    &$1603_{-656}^{+865}$    &CT &7\\
NGC 2655       &0.004670 &0301650301 &;                 &40.80&39.48  &$<$39.91                  &6.40*               &44*                 &$<$919                    &S2 &8\\
NGC 2992       &0.007710 &0147920301 &m1,m2,pn;         &42.93&41.15  &$40.82_{-0.14}^{+0.11}$ &$6.40_{-0.02}^{+0.02}$  &$35.3_{-34.8}^{+31.1}$  &$51_{-13}^{+15}$        &S2 &9\\
NGC 3079       &0.003723 &0110930201 &;                 &40.09 &39.67  &$39.09_{-0.18}^{+0.17}$ &$6.50_{-0.05}^{+0.05}$  &$111.4_{-45.0}^{+58.5}$ &$695_{-238}^{+342}$     &CT &10\\
               &         &0147760101 &;PN:25           &       &                        &                        &                        &                        &   &\\
NGC 3227       &0.003859 &0101040301 &;                 &41.49&40.27  &$39.91_{-0.07}^{+0.06}$ &$6.40_{-0.01}^{+0.01}$  &$43.0_{-43.0}^{+18.5}$  &$186_{-28}^{+28}$       &S1 &\\
NGC 3516       &0.008836 &0107460701 &;                 &42.46&40.99  &$40.51_{-0.13}^{+0.08}$ &$6.42_{-0.00}^{+0.00}$  &$<15.5     $    &$66_{-17}^{+12}$        &S1 &\\
NGC 3783       &0.009730 &0112210101 &m1,m2;            &42.92 &40.77  &$41.15_{-0.02}^{+0.02}$ &$6.40_{-0.00}^{+0.00}$  &$61.3_{-6.2}^{+5.7}$    &$108_{-5}^{+5}$         &S1 &\\
               &         &0112210201 &m2;              &       &                        &                        &                        &                        &   &\\
               &         &0112210501 &m2;              &       &                        &                        &                        &                        &   &\\
NGC 3786       &0.008933 &0204650301 &;                 &42.12&40.52  &$<$41.06                  &6.40*               &44*                 &$<$541                    &S2 &30\\
NGC 3982       &0.003699 &0204651201 &;                 &$<$39.77&39.55 &$<$39.02                  &6.40*               &44*                 &$<$1184                   &CT &12\\
NGC 4051       &0.002336 &0157560101 &;                 &40.89&39.50  &$39.25_{-0.07}^{+0.07}$ &$6.41_{-0.01}^{+0.01}$  &$57.5_{-20.7}^{+18.9}$  &$177_{-26}^{+31}$       &S1 &\\
NGC 4138       &0.002962 &0112551201 &;                 &41.08&38.92  &$39.13_{-0.34}^{+0.21}$ &$6.38_{-0.04}^{+0.05}$  &$<87.4     $    &$80_{-43}^{+48}$        &S2 &8\\
NGC 4151       &0.003319 &0112310101 &;                 &42.33 &40.70  &$40.61_{-0.02}^{+0.02}$ &$6.40_{-0.00}^{+0.00}$  &$48.5_{-2.5}^{+4.4}$    &$117_{-4}^{+4}$         &S1 &\\
               &         &0112830201 &m1,m2;           &       &                        &                        &                        &                        &   &\\
               &         &0112830501 &;                &       &                        &                        &                        &                        &   &\\
               &         &0143500101 &;                &       &                        &                        &                        &                        &   &\\
               &         &0143500201 &m1,m2;           &       &                        &                        &                        &                        &   &\\
               &         &0143500301 &m1,m2;           &       &                        &                        &                        &                        &   &\\
               &         &0112190201 &;                &       &                        &                        &                        &                        &   &\\
NGC 4168       &0.007388 &0112550501 &;                 &40.12&39.23  &$<$38.98                  &6.40*               &44*                 &$<$901                    &S2 &8\\
NGC 4235       &0.008039 &0204650201 &;                 &41.65&39.79  &$40.14_{-0.19}^{+0.15}$ &$6.41_{-0.03}^{+0.04}$  &$51.2_{-51.0}^{+49.7}$  &$284_{-98}^{+113}$      &S1 &\\
NGC 4258       &0.001494 &0110920101 &;                 &40.54 &38.57  &$38.24_{-0.36}^{+0.20}$ &$6.40_{-0.06}^{+0.03}$  &$<98.8     $    &$34_{-19}^{+20}$        &S2 &8\\
               &         &0203270201 &;PN:25           &       &                        &                        &                        &                        &   &\\
               &         &0059140101 &;                &       &                        &                        &                        &                        &   &\\
               &         &0059140201 &;                &       &                        &                        &                        &                        &   &\\
               &         &0059140401 &;                &       &                        &                        &                        &                        &   &\\
               &         &0059140901 &;                &       &                        &                        &                        &                        &   &\\
NGC 4378       &0.008536 &0301650801 &;                 &$<$40.54&39.47 &$<$39.45                  &6.40*               &44*                 &$<$1110                   &S2 &8\\
NGC 4388       &0.008419 &0110930701 &;                 &42.55&41.61  &$41.00_{-0.12}^{+0.11}$ &$6.43_{-0.02}^{+0.02}$  &$42.6_{-42.6}^{+27.6}$  &$156_{-38}^{+43}$       &S2 &13\\
NGC 4395       &0.001064 &0112521901 &;                 &40.18 &38.02  &$38.20_{-0.13}^{+0.11}$ &$6.37_{-0.03}^{+0.02}$  &$74.9_{-54.7}^{+38.7}$  &$87_{-23}^{+24}$        &S2 &8\\
               &         &0112522701 &;                &       &                        &                        &                        &                        &   &\\
               &         &0142830101 &;                &       &                        &                        &                        &                        &   &\\
               &         &0200340101 &;                &       &                        &                        &                        &                        &   &\\
NGC 4472       &0.003326 &0112550601 &;PN:25            &39.91 &$<$39.21 &$<$38.17                  &6.40*               &44*                 &$<$229                    &S2 &8\\
               &         &0200130101 &;                &       &                        &                        &                        &                        &   &\\
NGC 4477       &0.004520 &0112552101 &;                 &39.64&38.88  &$<$38.81                  &6.40*               &44*                 &$<$1436                   &S2 &8\\
NGC 4501       &0.007609 &0106060601 &;                 &40.13 &39.71  &$<$38.89                  &6.40*               &44*                 &$<$871                    &CT &14\\
               &         &0112550801 &;M1M2PN:25       &       &                        &                        &                        &                        &   &\\
NGC 4507       &0.011801 &0006220201 &;                 &42.61&41.01  &$41.13_{-0.06}^{+0.05}$ &$6.39_{-0.01}^{+0.01}$  &$48.2_{-16.9}^{+12.2}$  &$176_{-21}^{+22}$       &S2 &15\\
NGC 4565       &0.004103 &0112550301 &;M1M2PN:25        &39.96&38.89  &$<$38.65                  &6.40*               &44*                 &$<$386                    &S2 &8\\
NGC 4579       &0.005067 &0112840101 &;                 &41.39&39.21  &$39.57_{-0.26}^{+0.17}$ &$6.39_{-0.04}^{+0.04}$  &$0.7_{-0.7}^{+86.3}$    &$137_{-61}^{+66}$       &S2 &16\\
NGC 4593       &0.009000 &0109970101 &m2;               &42.65&40.38  &$40.91_{-0.11}^{+0.10}$ &$6.42_{-0.02}^{+0.01}$  &$0.1_{-0.2}^{+41.6}$    &$106_{-23}^{+27}$       &S1 &\\
NGC 4594       &0.003416 &0084030101 &;                 &40.63&38.83  &$<$38.99                  &6.40*               &44*                 &$<$243                    &S2 &17\\
NGC 4639       &0.003395 &0112551001 &;                 &40.18&38.59  &$<$39.08                  &6.40*               &44*                 &$<$871                    &S1 &\\
NGC 4698       &0.003342 &0112551101 &;                 &39.19&38.70  &$<$38.09                  &6.40*               &44*                 &$<$834                    &S2 &8\\
NGC 4939       &0.010374 &0032141201 &;                 &42.02&41.01  &$<$40.63                  &6.40*               &44*                 &$<$248                    &S2 &18\\
NGC 4945       &0.001878 &0112310301 &;M1M2PN:25        &40.19 &39.37  &$39.32_{-0.03}^{+0.04}$ &$6.40_{-0.00}^{+0.00}$  &$<4.4     $     &$696_{-52}^{+59}$       &CT &2\\
               &         &0204870101 &;M1M2PN:25       &       &                        &                        &                        &                        &   &\\
NGC 4968       &0.009863 &0200660201 &;                 &40.86 &40.81  &$40.12_{-0.14}^{+0.13}$ &$6.42_{-0.03}^{+0.02}$  &$<65.2     $    &$1252_{-338}^{+430}$    &CT &25\\
               &         &0002940101 &;                &       &                        &                        &                        &                        &   &\\
NGC 5005       &0.003156 &0110930501 &;                 &39.98&38.64  &$<$38.63                  &6.40*               &44*                 &$<$431                    &S2 &19\\
NGC 5033       &0.002919 &0094360501 &;                 &40.97&39.48  &$39.31_{-0.18}^{+0.13}$ &$6.42_{-0.03}^{+0.02}$  &$<67.4$    &$206_{-68}^{+74}$       &S1 &\\
NGC 5128       &0.001825 &0093650201 &m1,m2,pn;         &41.94 &39.86  &$39.85_{-0.07}^{+0.07}$ &$6.42_{-0.01}^{+0.01}$  &$30.7_{-30.7}^{+17.0}$  &$50_{-7}^{+8}$          &S2 &20\\
               &         &0093650301 &m2,pn;           &       &                        &                        &                        &                        &   &\\
NGC 5194       &0.001544 &0112840201 &;                 &39.39&39.11  &$38.43_{-0.15}^{+0.12}$ &$6.43_{-0.02}^{+0.02}$  &$37.2_{-38.1}^{+37.3}$  &$703_{-202}^{+229}$     &CT &21\\
NGC 5256       &0.027863 &0055990501 &;                 &41.81&42.04  &$40.57_{-0.22}^{+0.17}$ &$6.46_{-0.03}^{+0.03}$  &$<85.4$    &$418_{-165}^{+206}$     &CT &25\\
NGC 5273       &0.003549 &0112551701 &;                 &41.35&39.01  &$39.68_{-0.27}^{+0.14}$ &$6.43_{-0.03}^{+0.03}$  &$70.5_{-70.3}^{+50.0}$  &$212_{-97}^{+83}$       &S1 &\\
NGC 5506       &0.006181 &0013140101 &;                 &42.80 &41.27  &$40.56_{-0.08}^{+0.08}$ &$6.40_{-0.01}^{+0.01}$  &$20.4_{-20.4}^{+19.6}$  &$40_{-6}^{+7}$          &S2 &22\\
               &         &0013140201 &m1,m2;           &       &                        &                        &                        &                        &   &\\
               &         &0201830201 &;                &       &                        &                        &                        &                        &   &\\
               &         &0201830301 &;                &       &                        &                        &                        &                        &   &\\
               &         &0201830401 &;                &       &                        &                        &                        &                        &   &\\
               &         &0201830501 &;                &       &                        &                        &                        &                        &   &\\
NGC 5548       &0.017175 &0109960101 &m2;               &43.45 &41.00  &$41.32_{-0.05}^{+0.06}$ &$6.41_{-0.01}^{+0.01}$  &$68.4_{-18.5}^{+17.4}$  &$66_{-7}^{+9}$          &S1 &\\
               &         &0089960301 &;                &       &                        &                        &                        &                        &   &\\
               &         &0089960401 &;                &       &                        &                        &                        &                        &   &\\
NGC 5643       &0.003999 &0140950101 &;M1M2PN:25        &40.48&40.46  &$39.70_{-0.10}^{+0.09}$ &$6.41_{-0.02}^{+0.01}$  &$<48.9     $    &$1139_{-240}^{+275}$    &CT &6\\
NGC 6240       &0.024480 &0101640101 &;                 &42.51 &41.82  &$41.26_{-0.07}^{+0.07}$ &$6.41_{-0.01}^{+0.02}$  &$<40.6$    &$346_{-53}^{+57}$       &CT &32\\
               &         &0101640601 &;                &       &                        &                        &                        &                        &   &\\
               &         &0147420201 &;                &       &                        &                        &                        &                        &   &\\
               &         &0147420401 &;                &       &                        &                        &                        &                        &   &\\
               &         &0147420501 &;                &       &                        &                        &                        &                        &   &\\
               &         &0147420601 &;                &       &                        &                        &                        &                        &   &\\
NGC 7172       &0.008683 &0147920601 &;                 &42.58 &40.91  &$40.68_{-0.09}^{+0.06}$ &$6.42_{-0.01}^{+0.01}$  &$65.4_{-20.4}^{+19.2}$  &$88_{-16}^{+13}$        &S2 &18\\
               &         &0202860101 &;                &       &                        &                        &                        &                        &   &\\
NGC 7213       &0.005839 &0111810101 &m2;               &42.20&39.20  &$40.13_{-0.07}^{+0.07}$ &$6.42_{-0.01}^{+0.01}$  &$8.3_{-8.3}^{+36.8}$    &$74_{-10}^{+13}$        &S1 &\\
NGC 7314       &0.004763 &0111790101 &m1,m2;            &42.23&40.39  &$<$40.17                  &6.40*               &44*                 &$<$64                     &S2 &23\\
NGC 7469       &0.016317 &0112170101 &m1,m2;            &43.13 &41.34  &$41.24_{-0.09}^{+0.07}$ &$6.41_{-0.02}^{+0.02}$  &$64.1_{-29.1}^{+21.9}$  &$99_{-18}^{+18}$        &S1 &\\
               &         &0112170301 &m1,m2;           &       &                        &                        &                        &                        &   &\\
NGC 7479       &0.007942 &0025541001 &;                 &$<$40.55&40.57 &$39.81_{-0.46}^{+0.20}$ &$6.45_{-0.04}^{+0.04}$  &$78.6_{-78.6}^{+63.2}$  &$1016_{-666}^{+604}$    &S2 &8\\
NGC 7582       &0.005254 &0112310201 &;                 &41.27 &41.13  &$40.05_{-0.05}^{+0.04}$ &$6.41_{-0.01}^{+0.01}$  &$29.1_{-29.1}^{+13.4}$  &$343_{-33}^{+31}$       &CT &2\\
               &         &0204610101 &;                &       &                        &                        &                        &                        &   &\\
NGC 7590       &0.005255 &0112310201 &;                 &40.12 &39.62  &$<$38.98                  &6.40*               &44*                 &$<$632                    &CT &24\\
               &         &0204610101 &;                &       &                        &                        &                        &                        &   &\\
NGC 7603       &0.029524 &0066950301 &;                 &43.59 &41.15  &$<$42.06                  &6.40*               &44*                 &$<$302                    &S1 &\\
               &         &0066950401 &;                &       &                        &                        &                        &                        &   &\\
NGC 7674       &0.028924 &0200660101 &;                 &42.20&41.95  &$<$41.20                  &6.40*               &44*                 &$<$531                    &CT &33\\
UGC 12138      &0.024974 &0103860301 &;M1:25,M2:35      &43.13&41.20  &$<$41.78                  &6.40*               &44*                 &$<$515                    &S2 &19\\
\enddata
\tablecomments{
Column(1): Name of object.
Column(2): redshift.
Column(3): XMM-Newton observation ID.
Column(4): before the semicolon: which of the 3 EPIC detectors' exposures are suffering significant pileup. After the semicolon: in which of the 3 EPIC detectors' exposures, source region radius are reduced, instead of being the default 40 arcsec.
M1 and M2 are abbreviations for MOS1 and MOS2. The number after the colon represents the extraction region radius in arcsec.
Column(5-6): 2-10 keV and [OIV] luminosities in erg/s.
Column(7-10): luminosity, central energy, $\sigma$ and equivalent width of the narrow Fe K$\alpha$ line in rest-frame with $90\%$ errors. The upper limits to $\sigma$ are at 90\% confidence level.
$3\sigma$ upper limits of luminosity and equivalent width are given when the line is not detected.
* Denotes fixed quantities.
Column(11): Seyfert type. S1: Seyfert 1; S2: Compton-Thin Seyfert 2; CT: Compton-Thick Sy2.
Column(12): References for the X-ray classification of Sy2 galaxies (Compton-thin or Compton-thick).
}
\tablerefs{
(1) this work;
(2)\citealp{treister09};
(3)\citealp{churazov03};
(4)\citealp{risaliti09};
(5)\citealp{cardamone07};
(6)\citealp{maiolino98};
(7)\citealp{awaki08};
(8)\citealp{akylas09};
(9)\citealp{brenneman09};
(10)\citealp{iyomoto01};
(11)\citealp{madejski06};
(12)\citealp{shu07};
(13)\citealp{beckmann04};
(14)\citealp{brightman08};
(15)\citealp{matt04_4507};
(16)\citealp{dewangan04};
(17)\citealp{pellegrini03};
(18)\citealp{noguchi09};
(19)\citealp{gallo06};
(20)\citealp{rothschild06};
(21)\citealp{fukazawa01};
(22)\citealp{shinozaki06};
(23)\citealp{raymont02};
(24)\citealp{bassani99};
(25)\citealp{guainazzi05b};
(26)\citealp{shu08};
(27)\citealp{braito04};
(28)\citealp{balestra05};
(29)\citealp{greenhill08};
(30)\citealp{komossa97};
(31)\citealp{iwasawa01};
(32)\citealp{vignati99};
(33)\citealp{malaguti98};
}
\end{deluxetable}

\begin{deluxetable}{llccl}
\tabletypesize{\scriptsize}
\tablecaption{\label{addgauss}Additional Gauss lines}
\tablewidth{0pt}
\tablehead{
\colhead{Name}	&\colhead{Energy}	&\colhead{$\sigma$}	&\colhead{Flux}	&\colhead{References}\\
&\colhead{$(keV)$}	&\colhead{$(eV)$}	&\colhead{$(10^{-6} photons/cm^{-2}/s^{-1})$} &
}
\startdata
3C 120	&$6.96_{-0.03}^{+0.03}$ &$<72$ &$9_{-3}^{+5}$	&\citet{ballantyne_3c120}\\	
Circinus	&$7.04_{-0.01}^{+0.01}$ &$39_{-7}^{+12}$ &$41_{-2}^{+2}$	&\citet{molendi03_cir}\\
		&$7.47_{-0.01}^{+0.01}$ &$<31$ &$11_{-1}^{+2}$	&\\
		&$6.69_{-0.01}^{+0.01}$ &$67_{-14}^{+14}$ &$26_{-3}^{+3}$	&\\
I ZW 1	&$6.83_{-0.10}^{+0.11}$ &$312_{-115}^{+128}$ &$8_{-1}^{+3}$	&\citet{gallo07_izw1}\\
MCG-6-30-15	&$5.97_{-0.05}^{+0.09}$ &$310_{-102}^{+98}$ &$24_{-5}^{+11}$	&\citet{fabian03_mcg-6-30}\\
		&$6.89_{-0.02}^{+0.01}$ &$<27$ &$6_{-1}^{+1}$	&\\
MRK 335	&$7.02_{-0.04}^{+0.03}$ &$<58$ &$5_{-2}^{+3}$	&\citet{gondoin02}\\
NGC 1068	&$6.69_{-0.01}^{+0.01}$ &$63_{-14}^{+14}$ &$33_{-3}^{+5}$	&\citet{pounds06_1068}\\
		&$7.01_{-0.02}^{+0.02}$ &$68_{-23}^{+18}$ &$15_{-3}^{+3}$	&\\
NGC 1275	&$6.69_{-0.01}^{+0.01}$ &$<16$ &$51_{-5}^{+7}$	&\citet{churazov03}\\
		&$6.98_{-0.03}^{+0.03}$ &$<58$ &$5_{-3}^{+3}$	&\\
NGC 1365	&$5.78_{-0.06}^{+0.09}$ &$435_{-86}^{+55}$ &$67_{-24}^{+16}$	&\citet{risaliti09}\\
NGC 3516	&$6.42_{-0.01}^{+0.02}$ &$127_{-23}^{+14}$ &$36_{-8}^{+3}$	&\citet{turner02_ngc3516}\\
NGC 3783	&$7.03_{-0.02}^{+0.02}$ &$67_{-25}^{+22}$ &$16_{-3}^{+3}$	&\citet{reeves04_3783}\\
NGC 4151	&$7.08_{-0.03}^{+0.02}$ &$<65$ &$5_{-3}^{+4}$	&\citet{schurch_4151}\\
NGC 4945	&$6.68_{-0.03}^{+0.03}$ &$218_{-19}^{+26}$ &$18_{-2}^{+2}$	&\citet{schurch02_4945}\\
NGC 5506	&$6.53_{-0.06}^{+0.05}$ &$273_{-40}^{+37}$ &$85_{-18}^{+19}$	&\citet{matt01_5506}\\
		&$6.99_{-0.02}^{+0.03}$ &$<73$ &$11_{-3}^{+6}$	&\\
NGC 6240	&$6.67_{-0.02}^{+0.02}$ &$<54$ &$7_{-1}^{+2}$	&\citet{netzer05_6240}\\
NGC 7314	&$6.68_{-0.12}^{+0.21}$ &$360_{-111}^{+97}$ &$45_{-16}^{+26}$	&\citet{yaqoob03_7314}\\
\enddata
\tablecomments{
Column(1): Name.
Column(2-4): Centroid energy, $\sigma$ and flux of the line. The errors and upper limits are in 90\% confidence level.
Column(5): References in which the detections of the additional lines were also reported. 
}
\end{deluxetable}

\newpage

\end{document}